\documentstyle[preprint,prd,aps,floats]{revtex}
\newbox\rotbox
\newcommand{\be}{\begin{eqnarray}}
\newcommand{\ee}{\end{eqnarray}}
\newcommand\tag{\hbox to hsize}

\def\mytoday#1{{}\ifcase\month\or
January\or February\or March\or April\or May\or June\or
July\or August\or September\or October\or November\or December\fi
 \space \number\year}

\begin{document}
\draft
\noindent Submitted to {\it Physics Letters B} \hfill DOE/ER/40762--044

\noindent{\it }\hfill U of MD PP\#95--009

\noindent{\it }\hfill SUNY-NTG 94-52
\vskip .2 cm

\title{Chiral Phase Structure of Hot and Dense Nuclear Matter}

\author{Hilmar Forkel\thanks{Present address: European
Centre for Theoretical Studies in Nuclear Physics and Related Areas,
Villa Tambosi, Strada delle Tabarelle 286, I-38050 Villazzano, Italy}}
\address{Center for Theoretical Physics and Department of Physics, }
\address{University of Maryland, College Park, Maryland 20742-4111
(U.S.A.)}
\vskip -0.9 cm
\author{and}
\vskip -0.7 cm
\author{A. D. Jackson}
\address{Department of Physics, State University of  New York }
\address{at Stony Brook, Stony Brook, New York 11794 (U.S.A.)}
\date{November 1994}
\maketitle
\vskip -.7cm
\begin{abstract}
The description of nuclear matter bulk properties in the framework
of quark-based hadron models in curved unit cells is extended to
finite temperature. We discuss the analogy and interplay of density
and temperature effects in this formulation and study, as an
application, the temperature and density dependence of chiral
vacuum properties in the Nambu-Jona-Lasinio (NJL) model.
\end{abstract}

%\thanks{${^\dagger}$Supported in part by the US Department
% of Energy under Grant No.}

\narrowtext

The complexity of the strong interactions resides to a large part
in their ground state, the QCD vacuum, which is the focus of various
research directions in nuclear and particle physics. One of the
currently followed strategies towards uncovering the vacuum structure
is to study its response to changes in external conditions.
Systems with a large baryo-chemical potential and/or a high
temperature in sufficiently extended volumes of space-time are
well suited for this purpose and experimentally accessible.

Indeed, the above conditions are realized in the highly compressed
matter of dense stars, and they can be produced in relativistic
heavy-ion collisions. In both situations the baryon density and
temperature can become large enough to alter significantly the
structure of the matter ground state, which is expected to eventually
undergo transitions to new phases with modified symmetry properties.

In addition, hadron structure becomes a relevant factor for the
matter dynamics at high densities. Mainly as a consequence of
the changing vacuum properties, it also undergoes density and
temperature induced modifications, which have been estimated at the
mean-field level in various hadron models
\cite{modstudies,skyfind,njlfind}. Some of these studies explored
a new description of
nuclear matter bulk properties at zero temperature, which divides
the dense environment into curved cells of unit baryon number
\cite{HypSp1,HypSp2,for94}. Since high
baryon densities, in particular if produced in relativistic
heavy-ion collisions, are typically accompanied by temperatures of
the order of the QCD scale, this approach should be extended to
finite temperature, and this is the purpose of the present letter.

The crucial feature of the new approach, which generalizes the
standard unit cell descriptions for homogeneous or periodic media,
is the form of the unit cell, an intrinsically curved three-dimensional
hypersphere $S^3(L)$ of radius $L$. The hyperspherical cell provides
a new setting for the study of finite-density dynamics in hadron
models, which is in some aspects complementary to more conventional
approaches. In addition, it requires much less computational effort
than traditional unit-cell techniques, but reproduces results of
conventional dense-matter calculations in the Skyrme and
Nambu-Jona-Lasinio (NJL) models \cite{skyfind,njlfind} almost
quantitatively.

The recent generalization of the hypersphere approach to quark-based
hadron models \cite{for94} allows us to perform our study in the
standard Nambu-Jona-Lasinio model \cite{NJL69}. Our focus will be
mainly on the combined temperature and density dependence of chiral
vacuum properties and of the chiral phase transition. The pertinent
observables are well studied in the hypersphere approach at zero
temperature \cite{HypSp1,HypSp2,for94} and are expected to be rather
strongly $T$ dependent, as indicated, for example, by lattice
predictions of $T_c \simeq 140-160 \, {\rm MeV}$ for the chiral
restoration temperature \cite{kog91}.

In a hyperspherical unit cell $S^3(L)$ of radius $L$, the standard
NJL lagrangian takes the form \cite{for94}
\be
{\cal L} = \frac{i}{2} \left[\, \overline{q} \gamma^a ( e_a +
\omega_a) q -
\overline{q}( \overleftarrow{e}_a - \omega_a) \gamma^a q \, \right]
\,\, - \, m_o \, \overline{q} q \, + \, g \, (\overline{q} \,
\Gamma_a  q)\, (\overline{q} \, \Gamma^a q). \label{njlhyp}
\ee
The $q$'s are the (color triplet and iso-doublet) quark fields
with current mass $m_o$, and their characteristic four-quark contact
interaction is written in terms of the matrices $\Gamma^o =
1_{\gamma} 1_{\tau}, \,\,\,
\Gamma^i =  i \gamma_5 \tau_i \,\,\,(i=1,2,3)$. (The index $a \in
\{0,1,2,3\}$ is implicitly summed over in the interaction lagrangian
above.). In the unit cell context, the geometrical and topological
properties of $S^3 $ are mediating interactions with the dense
environment by characteristically modifying the NJL dynamics: the
flat-space gradients are replaced by their $S^3$
analogues $e_a$ (the arrow indicates operation to the left) and a
density-induced, isoscalar gauge field $\omega_a$, the spin connection,
emerges. Our notation follows ref. \cite{for94}, to which we also refer
for a detailed discussion of the new features of hyperspherical cells
in the context of quark-based hadron models.

For the study of the model (\ref{njlhyp}) at finite temperature we
adopt the standard imaginary time formalism \cite{kap89}. The
corresponding compactification of the time direction does not interfere
with the curvature and topology of the hypersphere, which only affects
the three spatial dimensions. In fact, the descriptions of temperature
and density in hyperspherical cells do not only coexist naturally
but also show direct analogies, which we will discuss further below.

In order to calculate the thermal Gibbs average of any desired
Green's function in the given framework, we analytically continue
to euclidean times $\tau = i x_0$ and impose the Kubo-Martin-Schwinger
(KMS) boundary conditions \cite{kub57}. For the quark propagator,
in particular, they require $S_\beta (\tau + \beta) = - \,
{\rm e}^{ \,\beta \zeta} \, S_\beta (\tau)$, where $\beta$ is the
inverse temperature and $\zeta$ is the chemical
potential which sets the average baryon number in the hyperspherical
unit cell to one (see below). The antiperiodicity in euclidean time
is a common property of all (fermionic) finite-temperature Green's
functions and implies that the model is now defined on a fully compact
and boundary-less domain, $S^3(L) \times S^1 (\frac{\beta}{2
\pi})$.

The NJL quark propagator in Hartree-Fock approximation has a
space-time independent quark self-energy with both scalar and vector
parts, $\Sigma = \Sigma_s - \gamma_4 \Sigma_v$. Its hypersphere
version was derived in ref. \cite{for94} at $T=0$ and can be easily
generalized to finite temperature:
\be
S_\beta ({\bf x}, \tau)  = {\rm e}^{ i ( \vec{ \Sigma } \hat{r}
\frac{\mu}{2} + i \Sigma_v \tau)} \, [\, \gamma_4 \,S_0 (\mu,\tau)  + i
\, \hat{r} \vec{\gamma} \, S_1
(\mu,\tau)  - S_2 (\mu,\tau) \, ]. \label{fullprop}
\ee
The spatial coordinates of the hypersphere are denoted in bold face,
{\it i.e.} ${\bf x}= \{\mu, \theta, \phi \}$, $\hat{r}=\hat{r}( \theta,
\phi)$ is the unit-vector in $R^3$, $\vec{ \Sigma }$ is the Dirac
spin matrix and the euclidean $\gamma$ matrices are chosen hermitean,
with $\{\gamma_{\mu}, \gamma_{\nu} \} = 2 \delta_{\mu \nu}$. The three
amplitudes $S_i (\mu,\tau)$ follow from their zero-temperature
counterparts \cite{for94} by Wick-rotating to euclidean time and
imposing the KMS condition. In the following we
will only need two of them explicitly, namely,
\be
S_2 (\mu,\tau) = \frac{ i m}{2 \beta V}  \sum_{n=1}^{ \infty} s_n (\mu)
\, \Theta_{\epsilon} (\Lambda - k_n) \sum_{m=- \infty}^{ \infty}
\frac{ {\rm e}^{- i \tilde{k}^{(m)}_4 \tau }}{(\tilde{k}^{(m)}_4)^2 +
\omega_n^2}   \label{S2spec}
\ee
and $S_0$, which is related to $S_2$ by a time derivative, $S_0
(\mu,\tau) = m^{-1} \, \partial_\tau S_2 (\mu,\tau)$.

These amplitudes represent the propagator in terms of
the discrete constituent quark spectrum in the unit cell, which is
completely determined by the geometrical properties of $S^3 (L)$ and
summarized in the generalized momenta $k_n=(2n+1)/(2L)$, energies
$\omega_n= \sqrt{k_n^2 + m^2}$ and degeneracies (apart from flavor
and color) $D_n = 2n(n+1)$.

Analogously, the compactified euclidean time dimension leads to a
discretized Matsubara frequency spectrum $k^{(m)}_4$ of the quarks,
which is additionally shifted by the vector part of the self-energy
and by the chemical potential:
\be
\tilde{k}^{(m)}_4 = k^{(m)}_4 - i (\zeta + \Sigma_v)\, , \qquad
k^{(m)}_4 = (2n+1) \frac{\pi}{\beta}.
\ee
Anticipating the well-known divergencies of quark loops in the NJL
model, we also introduced a soft momentum cutoff $\Lambda$ in eq.
(\ref{S2spec}). It has the form of a regularized theta function,
$\Theta_{\epsilon} (\Lambda - k) = ({\rm e}^{\epsilon L (k - \Lambda)}
+ 1)^{-1}$, with $\epsilon$ parametrizing the width of the transition
region.

The entire spatial coordinate dependence of the amplitudes $S_0$
and $S_2$ is contained in the set of functions
\be
s_n(\mu) = \frac{2 k_n L \sin (k_n L \mu) - \cos (k_n L \mu)
\tan \frac{\mu}{2} }{\sin \mu}, \quad \quad  s_n(0) = D_n,
\ee
which were introduced in ref. \cite{for94}. Their index refers to
the contribution from the $n$-th quark level. The  $s_n$ are
orthogonal on $S^3$ and normalized to the corresponding degeneracies:
\be
\frac{1}{2 \pi} \int_0^{2 \pi} d \mu  \,\sin^2 \mu  \,\, s_n(\mu) \,
s_m(\mu) = \delta_{n m} D_n.  \label{ortho}
\ee

With the finite-temperature quark propagator at hand, we can now
derive the combined temperature and density dependence of the
fundamental chiral properties, {\it i.e.} of the constituent quark
mass, the chiral order parameter and the pion decay constant. Up to
the small current masses, the constituent quark mass of the NJL model
is dynamically generated as the scalar part of the quark self-energy
\cite{for94}
\be
\Sigma (x) \equiv \Sigma =  i g \,  \{  \, 2 \Gamma_a \,
{\rm tr} \, [S_\beta^- (x,x) \Gamma_a ]  - \Gamma_a \, [S_\beta^+ (x,x)
+ S_\beta^- (x,x) ] \, \Gamma_a \,  \} . \label{sol}
\ee
(The superscript $\pm$ of the propagators distinguishes the coincidence
limits from positive and negative times.) From eq. (\ref{sol}) we
derive the scalar and vector components of the self-energy by
inserting the propagator (\ref{fullprop}),
\be
\Sigma_s &=& - 4  i g \, \left(1+2 n_c n_f \right) \, S_2 (0)
\nonumber \\ &=& \frac{2 m g}{ \beta V} (1+2 n_c n_f)  \sum_{n=1}^{\infty}
D_n  \Theta_{\epsilon} (\Lambda -  k_n ) \sum_{m=- \infty}^{ \infty}
\frac{1}{(\tilde{k}^{(m)}_4)^2 + \omega_n^2}, \label{sself} \\
\Sigma_v &=&  8  i g \,  S_0(0)
\nonumber \\ &=& \frac{4 i g}{
\beta V} \sum_{n=1}^{\infty} D_n  \Theta_{\epsilon} (\Lambda -
k_n ) \sum_{m=- \infty}^{ \infty}
\frac{\tilde{k}^{(m)}_4}{(\tilde{k}^{(m)}_4)^2 + \omega_n^2}.
\label{vself}
\ee
(Here, $n_c$ and $n_f$ denote the number of quark colors and flavors,
respectively.)

The Matsubara sums in the above expressions can be rewritten as
contour integrals \cite{kap89} and evaluated analytically, with
the result
\be
\Sigma_s &=& \frac{m g}{  V} (1+2 n_c n_f)  \sum_{n=1}^{\infty}
\frac{D_n}{\omega_n}  \Theta_{\epsilon} (\Lambda -  k_n )
\{1 - n^{+} (k_n) - n^{-} (k_n) \} , \label{sselffinal} \\
\Sigma_v &=& - \frac{2 g}{ V} \sum_{n=1}^{\infty} D_n  \Theta_{\epsilon}
(\Lambda -  k_n ) \{ n^{-} (k_n) - n^{+} (k_n) \},
\label{vselffinal}
\ee
where $n^{\pm} (k_n) = ({\rm e}^{\beta \omega_n^{\pm}} +1)^{-1}$
are the Fermi occupation number distributions\footnote{Note
that the zero-temperature limit of eqs. (\ref{sselffinal},
\ref{vselffinal}) leads back to the expressions of ref. \cite{for94}
(with three valence quarks in the cell). The subsequent $L \rightarrow
\infty$ limit reproduces the familiar zero-density results of the NJL
model \cite{NJL69}, as shown in ref. \cite{for94}.} with $\omega_n^{\pm}
\equiv \omega_n \pm (\zeta + \Sigma_v)$.

As usual, the dynamical quark mass is
obtained as the nontrivial solution of the gap equation $m(\rho,T)
= m_o + \Sigma_s(\rho,T)$, which formalizes the Hartree-Fock
self-consistency condition by requiring the value of $m$ in the
quark loop of $\Sigma_s$ (which appears implicitly in the energies
$\omega_n$) to be equal to the dynamically generated constituent
quark mass. The temperature and density dependence of the solutions
to the gap equation will be discussed below.

{}From the trace of the Hartree-Fock quark propagator we directly
obtain the second chiral property of interest, namely the scalar
quark condensate
\be
<<\bar{q} q>> \,\, =  -  i\, {\rm tr}_{\gamma,c} \left[
S_\beta(0) \right]
= - \frac{n_c}{(1+2 n_c n_f )} \frac{\Sigma_s}{g },
\label{qcond}
\ee
which serves as the standard order parameter of the chiral phase
transition. (The double brackets denote the thermal average of the
operator they enclose, and the trace is over color and
spinor indices.) The quark number density, or equivalently
the vector quark ``condensate'', can be similarly expressed as
$<<q^{\dagger} q>> \, = - n_c \Sigma_v/(2 g)$. Since $<<q^{\dagger}
q>>$ is kept fixed in the unit cell at $3 \rho / 2 =  3/(2 V)$, the
above equation determines the chemical potential as a function of
the quark mass, the baryon density and the temperature, {\it i.e.}
$\zeta = \zeta(m,\rho,T)$.

The remaining chiral quantity of interest is the pion
decay constant $f_{\pi}$. We obtain it from the Bethe-Salpeter
amplitude $\chi^a (x,y;p_4)$ of the pion in its rest frame ({\it
i.e.} at ${\bf p} =0$), which is proportional to the one-pion
matrix element of the axial current and thus to $f_{\pi}$:
\be
<<0|A_{\mu}^a (x)| \pi^b (p_4) >> \,\,= -  \, {\rm tr}_{\gamma,c,f}
\, [ \gamma_{\mu}  \gamma_{5} \frac{\tau^a}{2} \chi^b (x,x;p_4) ]
\,\frac{ e^{-i p_4 \tau} }{\sqrt{2 \omega_{\bf p} V} }
= - p_4 f_{\pi} \delta^{a b} \frac{ e^{-i p_4
\tau} }{\sqrt{2 \omega_{\bf p} V} }. \label{axme2}
\ee
Due to the periodicity in the euclidean time direction, the pion
energy (which is later analytically continued back to the pion mass)
becomes a bosonic Matsubara frequency $p_4^{(l)} = 2 l \pi / \beta$.
The trace is over color, flavor and spinor indices and connects the
external quark lines of the Bethe-Salpeter amplitude to the axial
vector vertex.

The solution of the NJL Bethe-Salpeter equation for the pion
in ladder approximation is, apart from the external quark
propagators, just the pseudoscalar pion-quark Yukawa
vertex, with the coupling determined by the Goldberger-Treiman
relation at the quark level, $g_{\pi q q} =  m /f_{\pi}$.
Adapted to finite temperature it reads
\be
\chi^a (x,y;p_4) =  - \frac{ m }{f_{\pi}} \,\int d \mu(z)
e^{-i p_4 \tau_z} S_\beta(x,z) \, \tau^a \gamma_5 \, S_\beta(z,y)
\,\, + O(p_4^2) . \label{piwf}
\ee
The integration range in eq. (\ref{piwf}) is $S^3(L) \times S^1(
\frac{\beta}{2 \pi})$ with the measure  $d \mu(z) =   L^3 \sin^2
\mu \sin \theta \, d\mu  \, d \theta \, d \phi \, d \tau_z$.
{}From eq. (\ref{axme2}) we now obtain
\be
f^2_{\pi} = 4 i \pi n_c n_f \frac{L^3}{p_4} \int_0^{2 \pi} d \mu
\, \sin^2 \mu \int_{0}^{\beta} d \tau e^{- i p_4 \tau} \,
\partial_\tau \, \{ S_2 (\mu,\tau) \, S_2 (\mu,-\tau) \},
\ee
and after a partial integration in the $\tau$ integral (the surface
term vanishes, due to the periodicity properties of both $S_2$ and
the exponential)
we insert the explicit form (\ref{S2spec}) of the propagator
amplitude $S_2$. The integration over the geodesic hypersphere
coordinate $\mu$ is immediately performed by using the
orthogonality relation (\ref{ortho}), and the integral over $\tau$
combines the two independent Matsubara sums into one,
\be
\int_0^{2 \pi} &d\mu& \, \sin^2 \mu \int_{0}^{\beta} d \tau e^{-
i p_4^{(l)} \tau} \, S_2 (\mu,\tau) \, S_2 (\mu,-\tau) \nonumber \\
&=& - \frac{\pi}{2} \frac{m^2}{ \beta V^2}  \sum_{n=1}^{\infty}
D_n  \Theta_{\epsilon} (\Lambda -  k_n ) \sum_{m=- \infty}^{ \infty}
\frac{1}{(\tilde{k}^{(m)}_4 + \frac{p_4^{(l)}}{2})^2 + \omega_n^2}
\,\, \frac{1}{(\tilde{k}^{(m)}_4 - \frac{p_4^{(l)}}{2})^2 +
\omega_n^2} ,
\ee
which can again be evaluated explicitly in terms of contour integrals.
After continuing back to Minkowski energies and specializing to the
chiral limit, we arrive at our final expression for the square of
the pion decay constant:
\be
f^2_{\pi} &=& \frac{n_c n_f }{4} \frac{m^2}{ V} \sum_{n=1}^{\infty}
\frac{D_n}{\omega_n^3} \, \Theta_{\epsilon} (\Lambda -  k_n )
\left\{1 - n^{+} (k_n) - n^{-} (k_n)  \right\}. \label{fpi}
\ee

Note that the explicit temperature dependence (the factor in curly
brackets) both of this expression and of the scalar quark self-energy
(\ref{sselffinal}) are identical. The overall $T$ dependence differs,
however, due to the implicit $T$ dependence of the $\omega_n$.
Nevertheless, the factor of $m^2$ ensures the decoupling of the pion
at the chiral phase transition.

The final expressions for the gap equation and the chiral vacuum
properties have the simple and
intuitive form which is typical for hypersphere results, despite the
perhaps somewhat unfamiliar formalism used in their derivation. In
fact, the density-induced modifications of the flat-space results
affect primarily the constituent quark spectrum and not the general
form of the expressions. This makes the transition to finite
temperature particularly straightforward in the NJL model, where
a constant mean field (the constituent quark mass) breaks chiral
symmetry. It can be ``heated'' as in standard perturbation theory,
since it is explicitly generated by loops. This is in contrast
to the Skyrme model, for example, where thermal fluctuations around
the spatially varying soliton field would have to be taken into
account.

To prepare the quantitative discussion of our results, we fix the
model parameters at the same values as in ref. \cite{for94}: $g=
4.08 \, {\rm GeV}^{-2}$, $\Lambda = 700 \, {\rm MeV}$ and $\epsilon
= 2.4$. The behavior of the chiral order parameter as a function of
the temperature is shown in Fig. 1 for different values of the
density. At small $T$ the condensate is only weakly temperature
dependent, even at rather high densities. With increasing temperature
it starts to drop rather steeply on the approach to the the critical
temperature $T_c$ of the chiral restoration transition. As one
would expect, $T_c$ itself decreases monotonically with density.

Figure 2 displays the temperature dependence of the pion decay
constant, again with the density as a parameter. As expected,
$f_\pi$ reaches zero at the same (density dependent) critical
temperature as the quark condensate, confirming that the pion
decouples from the vacuum when chiral symmetry is restored.
However, $f_\pi$ starts to decrease significantly at somewhat
lower temperatures than the quark condensate, exactly as was
observed in the conventional chemical potential approach
\cite{njlfind}.

Since the description of density and temperature in hyperspherical
cells is built on the compactification of space and time directions,
respectively, the explicit density and temperature dependence of the
chiral observables factorizes. The implicit coupling through the
solution $m(\rho,T)$ of the gap equation leads, nevertheless, to a
non-trivial interplay between temperature and density. It is
therefore remarkable that the temperature and density
dependence of the observables follows rather closely the one obtained
in the conventional treatment \cite{njlfind}, even though it can
be traced to specific features of the hyperspherical cell. This seems
to be a recurrent theme in hypersphere calculations, already
encountered in the studies of the Skyrme and NJL models at zero
temperature.

We conclude with several comments. Our study of the combined
temperature and density dependence of chiral properties in the
Nambu-Jona-Lasinio model is the first application of the hypersphere
approach at finite temperature. The extension to finite temperature
fits naturally into the hypersphere framework and extends its range
of applicability a step farther towards (if only qualitative) contact
with experiment.

Our results resemble rather closely those of the conventional
chemical potential approach to the NJL model, which approximates hot
and dense matter as a constituent quark gas. Some thermodynamical
quantities are sensitive to this neglect of the quark clustering
into color singlets \cite{kus90}, and the hypersphere description
does not resolve this problem, despite its inclusion of, on average,
three quarks into a unit cell. The latter does not lead to confinement
or clustering of the quarks, since the periodicity of the hypersphere
coordinates allows them to enter neighboring cells by travelling more
than once around the sphere, in complete analogy with the effect of
periodic boundary conditions in standard unit cells.

Our treatment of the temperature dependence in the unit cell is in
some sense minimal, since one could have allowed a temperature
dependence of the unit-cell metric in addition to the thermodynamics
of the quarks. It is conceivable,
for example, that an extended variational approach would find a
different cell shape favorable at high temperatures, leading to an
analog of a Landau transition in crystals. This is not likely,
however, since the cell metric has its origin mainly in the chiral
symmetry properties of the model {\it dynamics}, which do not change
at finite temperature.

Finally, it is interesting to note that the realization of finite
density and temperature are remarkably similar in hyperspherical
unit cells. In particular, {\it both} are described by compactifying
space-time directions, leading to discrete momenta and energies,
respectively, with their scale set by $\rho$ and $T$. (The time
direction stays flat, of course, since a one-dimensional circle
has no intrinsic curvature.)

Since there are many similarities in the effects of finite baryon
density and temperature, for example on phase transitions and on
hadron properties, it is tempting to speculate that a closely
analogous description of both would make their common aspects more
transparent and that the hypersphere approach gives a hint in this
direction.

This work was supported by the U.S. Department of Energy under
grants no. DE-FG02-93ER-40762 and DE-FG02-88ER-40388.

\newpage

\begin{figure}[bht]
\caption{ $-<<\bar{q} q>>^{1/3}$ as a function of temperature for
five values of the baryon density, $\rho / \rho_0 =$ $\{1.58, 1.34,
0.89, 0.46, 0.01 \}$, in units of the nuclear matter saturation
density  $\rho_0 = 0.17 \, {\rm fm}^{-3}$. (The uppermost curve
corresponds to the lowest density, and the ones below are arranged
in order of increasing density.) }
\label{fig1}
\end{figure}

\begin{figure}[hbt]
\caption{The pion decay constant as a function of temperature,
parametrized by the same five values of the density as in Fig. 1.}
\label{fig2}
\end{figure}


\newpage

\begin{thebibliography}{99}

\bibitem{modstudies} E. Shuryak, {\it The QCD Vacuum, Hadrons
and the Superdense Matter} (World Scientific Pub. Co., Singapore,
1988) and references therein

\bibitem{skyfind}  I. Klebanov, Nucl. Phys. {\bf B 262} (1985) 365;
L. Castillejo, P.S.J. Jones, A.D. Jackson, J.J.M. Verbaarschot and
A. Jackson, Nucl. Phys. {\bf A 501} (1989) 801

\bibitem{njlfind} T. Hatsuda and T. Kunihiro, Phys. Rev. Lett.
{\bf 55} (1985) 158;V. Bernard, U.-G. Meissner and I. Zahed, Phys.
Rev. {\bf D 36} (1987) 819; T. Hatsuda and T. Kunihiro, Phys. Rep.,
to be published, and references therein

\bibitem{HypSp1} N.S. Manton and P.J. Ruback, Phys. Lett. {\bf 181
B} (1986) 137; N.S. Manton, Commun. Math. Phys. {\bf 111} (1987) 469;
A.D. Jackson, NORDITA preprint 87-38 N (1987)

\bibitem{HypSp2}  A.D. Jackson, A. Wirzba and L.Castillejo, Phys.
Lett. {\bf 198 B} (1987) 315, Nucl. Phys. {\bf A486} (1988) 634;
H. Forkel, A.D. Jackson, M. Rho, C. Weiss, A. Wirzba and H. Bang,
Nucl. Phys. {\bf A 504} (1989) 818; A. Wirzba and H. Bang, Nucl.
Phys. {\bf A 515} (1990) 571 and references therein

\bibitem{for94} H. Forkel, Phys. Lett. {\bf 280 B} (1992) 5;
University of Maryland preprint \#94-037 (1994), to appear in
Nucl. Phys. {\bf A }

\bibitem{NJL69} Y. Nambu and G. Jona-Lasinio, Phys. Rev. {\bf 124}
(1961) 246, 255

\bibitem{kog91} J.B. Kogut, D.K. Sinclair and K.C. Wang,
Phys. Lett. {\bf 263 B} (1991) 101

\bibitem{kap89} J.I. Kapusta, {\it Finite-Temperature Field Theory},
Cambridge University Press, Cambridge (1989)

\bibitem{kub57} R. Kubo, J. Math. Soc. Japan {\bf 12} (1957) 570;
P.C. Martin and J. Schwinger, Phys. Rev. {\bf 115} (1959) 1342;
R. Haag, N.M. Hugenholz and M. Winnink, Commun. Math. Phys. {\bf 5}
(1967) 215

\bibitem{kus90} K. Kusaka, Phys. Lett. {\bf 269 B} (1991) 17


\end{thebibliography}
\end{document}